\documentclass[reprint, amsmath,amssymb, aps, prl, floatfix]{revtex4-2}

\usepackage{graphicx}
\graphicspath{{figures/}}
\usepackage{bm}
\usepackage{physics} 
\usepackage{xcolor}

\usepackage[range-phrase = --, print-unity-mantissa = false, exponent-product = \cdot, per-mode = single-symbol, uncertainty-mode = separate, inter-unit-product = .]{siunitx}

\DeclareSIUnit\litre{l}

\newlength{\maxfigwidth}
\newcommand{\mucl}{\mu_\text{cl}}
\newcommand{\vdrift}{v_\text{drift}}
\newcommand{\Qthres}{Q^*}

\usepackage[normalem]{ulem}
\newcommand{\revisemore}{}
\newcommand{\revisecrossmore}[1]{}

 \newcommand{\revise}{}
 \newcommand{\revisecross}[1]{}

\begin{document}

\title{%
Phase-locking parametric instability coupling longitudinal and transverse waves on rivulets in a Hele-Shaw cell}

\author{Grégoire Le Lay}
\email{gregoire.le-lay@u-paris.fr}
\author{Adrian Daerr}

\affiliation{%
 Matière et Systèmes Complexes UMR\,7057 CNRS, Université Paris Cité, 
 75231 Paris cedex 13, France%
}%

\date{\today}




\begin{abstract}
We report an instability exhibited by a fluid system when coupling two distinct types of waves, both linearly damped. While none of them is unstable on its own, they amplify one another resulting in a previously unreported convective instability. An external excitation is used to induce a parametric cross-coupling between longitudinal and transverse deformations of a liquid bridge between two vertical glass plates. Coherent amplification results for waves satisfying a synchronization condition, which selects a precise wavelength. We derive a model for this instability using depth-averaged Navier--Stokes equations, showing the physical origin of the co-amplification, and confirm its relevance experimentally.
%
%
%
%
%
%
Our findings open new perspectives in the study of parametrically controlled pattern formation, and invite to search for analogous parametric cross-coupling instabilities in other systems exhibiting distinct wave types, from plasma to elastic media.
\end{abstract}

\maketitle


For more than one and a half centuries, the study of hydrodynamic
instabilities has driven our understanding of dynamical systems, and
led to the development of tools to tackle non-linear systems with many
degrees of freedom that are used in all realms of physics and indeed
all sciences. Examples of hydrodynamic instabilities include the laminar-to-turbulent transition~\cite{avila2023pipeflow}, the Rayleigh--Plateau instability of a liquid cylinder~\cite{eggers2008} or the Kelvin--Helmholtz instability of the interface between fluid phases moving relative to one another, as in wind blowing over water giving rise to waves~\cite{rabaud2020}. Understanding these instabilities is of tremendous importance in environmental, biological and other natural settings as well as in many industrial processes, where the instability can be desirable as in combustion or printing, or deleterious as in coating~\cite{charru2011}.



This paper reports on an original parametric instability that should be relevant in many contexts outside hydrodynamics. Parametric instabilities arise from the temporal variation of a multiplicative parameter.  In hydrodynamics (Faraday instability of an accelerated liquid~\cite{douady1989}) or optics (parametric amplification of optical signals~\cite{1965opticalparametric}), this variation usually creates a nonlinear coupling of a wave-field to itself (surface height resp.\ electric field in the given examples).
In contrast, the instability described here involved the coupling of two distinct wave types, which do not interact in the absence of external forcing. The parametric coupling of distinct modes can also occur in plasma, e.g. Langmuir and ion acoustic waves, under the effect of an external dipolar field~\cite{nishikawa1968parametric,liu1986parametric}.
As a multitude of physical systems can sustain distinct wave types (e.g. compressive and shear waves in elastic media), we expect that analogous non-linear wave coupling instabilities may occur in other very different contexts from geophysics and acoustics to astrophysics.

Here we subject a liquid filament, henceforth termed rivulet, to homogeneous acoustic forcing, and describe a previously unreported instability where the path followed by the rivulet becomes sinuous, while simultaneously the streamwise mass distribution becomes inhomogeneous. We show that both features, although damped under normal conditions, grow by amplifying one another through a parametric coupling created by the the acoustic forcing. 
When the difference between fluid advection velocity and sinusoidal wave speed precisely matches the ratio of perturbation-wavelength and -period, the coupling becomes coherent and phase-locked, leading to reciprocal amplification. For the sake of clarity we stress that this cross-coupling and its parametric origin is the original finding of this article, and fundamentally distinguishes the resulting instability from the unforced, inertial meandering instability occuring above a threshold flow rate~\cite{drenkhan2007, daerr2011}.

\paragraph{Experimental setup} We inject liquid between two vertical and parallel glass plates separated by a gap of air
of thickness $b = \SI{0.6}{\milli\metre}$, forming a Hele-Shaw cell. The liquid (perfluorinated polyether PFPE, Galden HT135, density
$\rho = \SI{1.71}{\gram\per\milli\litre}$, surface tension
$\gamma = \SI{17}{\milli\newton\per\metre}$, kinematic viscosity
$\nu = \SI{1}{\milli\metre\squared\per\second}$) totally wets the glass.
The liquid forms a bridge joining the plates and falling downwards.
Since the plate separation $b$ is chosen inferior to the capillary
length, the bridge is in first approximation bounded by
semi-cylindrical interfaces meeting the glass with vanishing contact
angle~\cite{parkHomsy84}.

\revise{The liquid is injected into the cell through a pipette tip fed by a gear pump. Underneath the cell the fluid falls into a container that is continuously weighted, allowing us to measure the flow rate $Q$. 
}

\begin{figure}
    \centering
\includegraphics[width=0.8\maxfigwidth]{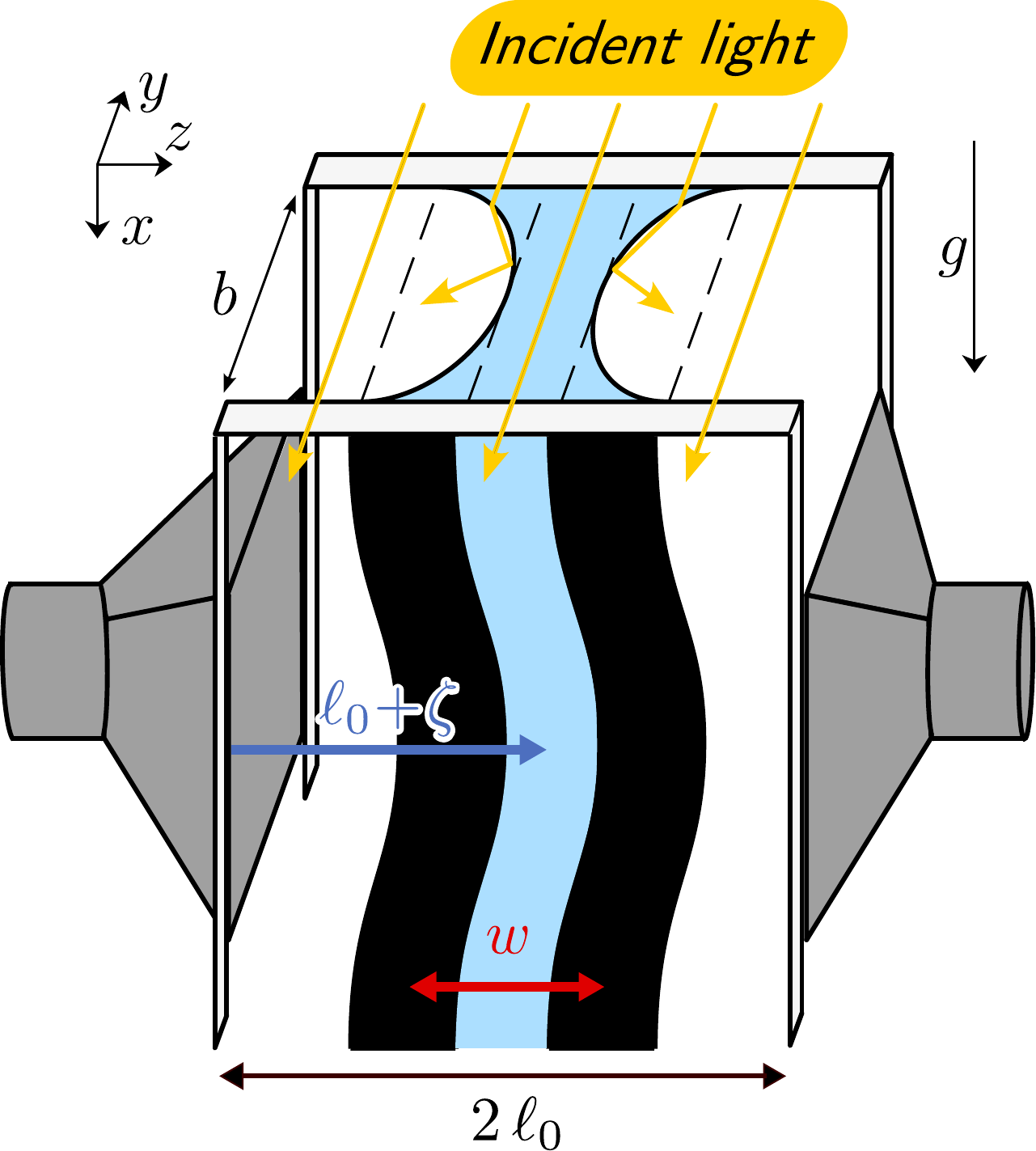}
\caption{Sketch of the experimental apparatus, not to scale. The glass
  panes are \SI{1}{\metre} high and \SI{30}{\centi\metre} wide, and
  set $b=\SI{0.6}{\milli\metre}$ apart. The lateral boundaries are
  closed except for a \SI{20}{\centi\metre} range, vertically
  centered, where loudspeakers impose the pressure. When viewed
  head-on, the rivulet appears delimited by dark bands where the
  curved interface refracts light away from the optical axis.}
\label{fig:setup}
\end{figure}

We use a camera to look at the rivulet which is back-lit with
quasi-collimated light. The two regions of curved menisci appear as dark
bands on a bright background, framing a central light band where light
passes unhindered through the bulk (fig.~\ref{fig:setup}). This allows
us to record the position of both the menisci as a function of $x$,
from which we define the rivulet position $\zeta(x, t)$ as the middle of
the bright region and the rivulet width $w(x, t)$ as the distance
between the two menisci. By measuring the width of the dark regions,
we are also able to know if the menisci are still semi-circular or
have been deformed.

When not subject to forcing, the rivulet flows straight vertically or
exhibits spontaneous meandering depending on whether the flow rate is
inferior or superior to a critical flow rate $\Qthres$, the origin of
which has been studied previously \cite{daerr2011}. In both cases, the
rivulet is always observed to be of constant width. Indeed, since the
curvature of the interface in the transverse \revise{$(y,z)$} plane is
fixed by the cell spacing, the rivulet is not subjected to the
Plateau--Rayleigh instability and any variation in width is linearly
damped.

The rivulet behaves as a 1-dimensional membrane effectively splitting
the cell in two regions, into which we force air using speakers on the
sides of the cell (fig.~\ref{fig:setup}) driven by a sinusoidal signal
of frequency $f_0$ in a push-pull configuration: when one speaker
pushes air into the space on the left of the rivulet, the other draws
air on the right, and the process is reversed half a period later.
Since the acoustic wavelength in air corresponding to the frequency
used is always larger than the cell width, the rivulet is subjected to
spatially homogeneous forcing over a region spanning
\SI{20}{\centi\metre} lengthwise. \revise{The neutral line around
  which the rivulet oscillates can display a small shift in the $z$
  direction, with respect to the path in the absence of forcing, over
  the scale of the excited portion of the cell. This is a consequence
  of a slight asymmetry in amplitude of the movements induced
  by the two speakers.}

\begin{figure}
    \centering
\includegraphics[width=\maxfigwidth]{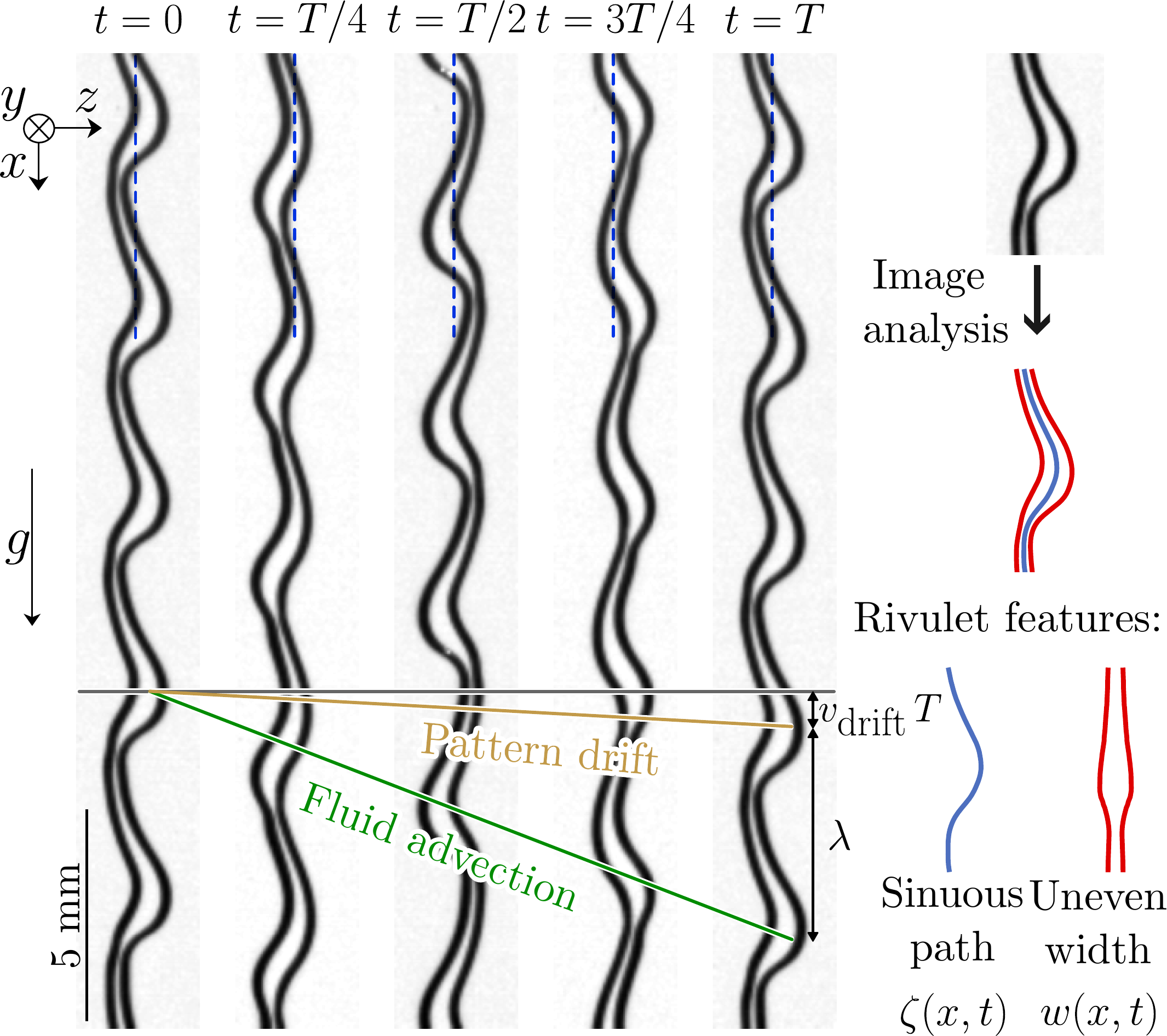}
\caption{Left: Snapshots of the rivulet over one forcing period \revise{$T\!=\!\SI{20}{\milli\second}$, for $Q\!>\!Q^*$ ($Q\!=\!\SI{46 \pm 1}{\cubic\milli\metre\per\second}$)}. The phase velocity $v_{\mathrm{drift}}$ of the sinuous deformation is smaller than that at which the liquid bulges flow downstream. Note how the pattern reproduces exactly after one period $T$, up to a translation. 
The dashed blue line represents the position of the rivulet averaged
in time and space. The spatially averaged (over $x$) rivulet position
coincides with this line at $t\!=\!0, T/2, T$ and the rivulet is
off-centered on the left at $t\!=\!T/4$ and symmetrically on the right at
$t\!=\!3T/4$. Right: decomposition of the rivulet profile into the $z$-wise
deformation of the center-line ($\zeta$, in blue) and the width modulation ($w$, in red).}
\label{fig:instability}
\end{figure}

\paragraph{Results}
At leading order, the rivulet responds to the forcing by moving
sideways, i.e.~along $z$, harmonically. At low to moderate frequencies
we observe that inertia is negligible, that is the rivulet
displacement is in phase with that of the loudspeaker membranes. This
sideways movement of the whole rivulet at the\revisecross{~excitation} \revise{forcing }frequency is
always present, as indicated by the displacement relative to the blue
dashed line in fig.~\ref{fig:instability}. \revisemore{In the experiments that follow, we used transverse displacements of \SIrange{0.2}{2}{\mm}, depending on the frequency.}

On top of this synchronous sideways movement, when the forcing
amplitude is \revisecrossmore{sufficient}\revisemore{above a certain threshold that depends on the forcing frequency}, the rivulet adopts a sinuous trajectory with
a well defined wavelength $\lambda$ that is orders of magnitude below the
acoustic wavelength. Moreover, the width of the rivulet is also
modulated with the same spatial periodicity
(fig.~\ref{fig:instability} and movie~S1\revise{; this modulation is sometimes
  termed \textsl{varicose mode} in the literature}). Movie~S2 shows
the initial growth after the onset of forcing, while movie~S3
shows how the perturbations quickly straighten out when the excitation is
switched off.

\begin{figure}
    \centering
\includegraphics[width=\maxfigwidth]{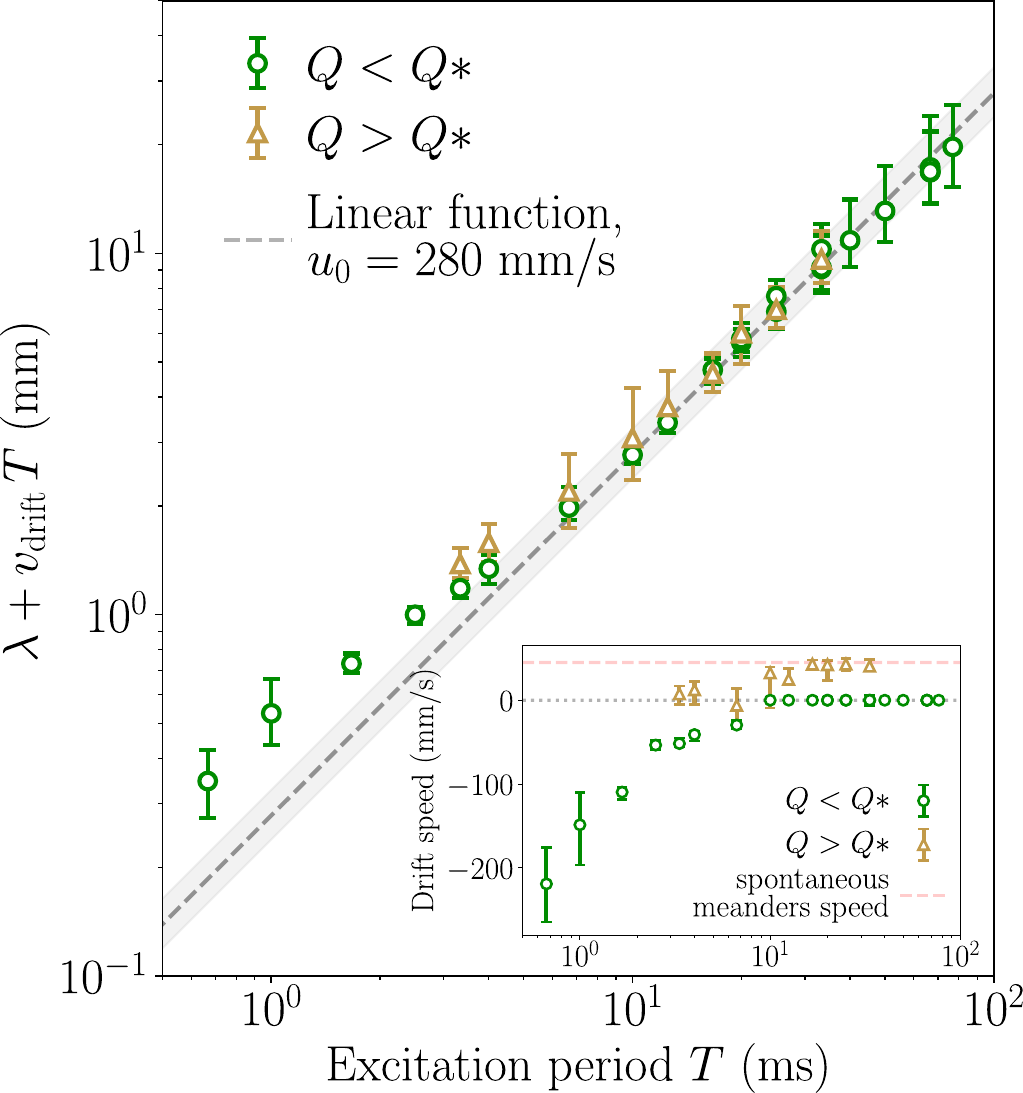}
\caption{The distance traveled over one period by the width modulation as a function of the \revisecross{excitation} \revise{forcing} period for two flow rates \revise{$Q$, \SI{26 \pm 1}{\cubic\milli\metre\per\second} and \SI{46 \pm 1}{\cubic\milli\metre\per\second}, being respectively below and above the spontaneous meandering threshold $\Qthres$}. The dashed line is the linear function of slope $u_0$. Inset: The sinuous pattern drift speed $\vdrift$ as a function of the \revisecross{excitation} \revise{forcing} period.}
\label{fig:dispersionrelation}
\end{figure}

The width modulations are advected streamwise at a velocity only
weakly depending on frequency and flow rate, whereas the sinuous
pattern is either static or drifts slowly at a \revise{drift }speed
$\vdrift$ that can be zero, positive or negative \revise{and that
  depends on the forcing frequency} and the flow rate
(fig.~\ref{fig:dispersionrelation} inset).

This reorganization of the rivulet is observed for a wide range of
frequencies (\SIrange{10}{1000}{\hertz}). The wavelength nears the
system size at the low end of this range, and drops below the gap size
$b$ and optical resolution at high frequencies, indicating that the
frequency range of observable response could be extended even further
through appropriate modifications of the set-up.

When the drift speed is zero, we observe that the rivulet profile is
exactly the same, both in lateral displacement and width, every period
$T$. This also holds, up to a translation in $x$, when the drift speed
is non-zero (fig.~\ref{fig:instability}). In other terms the path and
width modulations are phase-locked. This suggests that the wavelength
selection is given by the relative speed of the width modulation with
respect to the path modulation: after one period, compared to the path
modulation, the width modulation has traveled exactly one wavelength
further downstream. The wavelength thus acts as a degree of freedom
that allows the rivulet to respond to any\revisecross{~excitation} \revise{forcing }frequency (movie~S1).

This is remarkably confirmed by plotting the distance traveled by the
width modulation --- i.e. the wavelength \revise{$\lambda$} plus path drift
distance in one period \revise{$\vdrift\,T$} --- as a function of the
\revisecross{excitation} \revise{external forcing} period $T$
(fig.~\ref{fig:dispersionrelation}). \revise{Assuming the phase speed
  of the width modulations matches the bulk (Darcy) flow speed
  $u_0 = g\,b^2/(12\,\nu)=\SI{280 \pm 40}{\milli\metre\per\second}$, one
  expects the relation $\lambda+\vdrift\,T = u_0\,T$ to hold.} We find
quantitative agreement over two orders of magnitude with this
prediction, without any adjustable parameter.

The points are slightly offset from the curve for periods smaller than
\SI{4}{\milli\second}, an effect that we attribute to the meniscus
deformation. Indeed, the semi-circular shape of the meniscus is
maintained for slow rivulet movements, but viscous dissipation changes
the dynamic contact angle on the glass significantly for faster
motion. When the rivulet is subjected to fast transverse movements,
the interface flattens where the fluid advances while the concavity
increases where it retreats. For this reason, at high frequencies we
experimentally observe the interface to become non-circular, with the
meniscus shadows showing time oscillating asymmetry (not shown). We
thus expect the average viscous dissipation to differ from the
semi-cylindrical meniscus case, affecting the fluid velocity and/or
the sinuous drift velocity.

The measured drift speed of the path modulation
(fig.~\ref{fig:dispersionrelation} inset) shows a non-trivial
behavior: for flow rates below the spontaneous meandering threshold
($Q<\Qthres$) the sinuous pattern is stationary for frequency below
\SI{100}{\hertz}, and drifts upstream at higher frequencies. For flow
rates above $\Qthres$ and at low frequencies, the sinuous pattern
moves downstream at the spontaneous meanders' \revise{phase
}speed\revise{, which we measure independently in the absence of
  forcing}.


\paragraph{Discussion} In this section we propose a model
for the rivulet dynamics, based on the dominant physical ingredients.
We identify the mechanism for the unstable cross-amplification of
phase-locked path and width modulations.

The action of the speakers can be taken as equivalent to that of two infinite rigid vertical walls placed symmetrically at distance $\revise{\pm}\ell_0$ from the \revise{rest position of the straight }rivulet\revise{~at $\zeta = 0$}, moving horizontally so that their position relative to the situation at rest is given by $Z(t) = Z_0\cos(2\pi f_0 t)$, and acting like pistons on the air to the left and right of the rivulet. From Mariotte's law, for small displacements it follows that the force per unit length exerted on the rivulet is $b \rho \Pi(Z-\ev{\zeta}_x)$ with $\Pi = 2P_0/(\rho \ell_0)$, where $P_0$ is the atmospheric pressure and $\ev{\zeta}_x$ is the space-averaged position of the rivulet. Thus the rivulet behavior is given by
\begin{align}
\label{eq:ns}
    w(\partial_t + \beta\vb{u}\cdot&\grad) \vb{u}
    = w\vb{g} - w\mu \vb{u}
    +  w \Gamma \grad(\partial_{xx}w)\nonumber\\
    &+ \qty(\Gamma \partial_{xx}\zeta - \mucl \vb{u}\cdot\vb{n} + \Pi(Z-\ev{\zeta}_x) )\vb{n}, \\
\label{eq:massconservation}
    (\partial_t + \vb{u}\cdot&\grad) w = -w \grad \cdot \vb{u}.
\end{align}
The first equation is the depth-averaged Navier--Stokes equation integrated over the width of the rivulet $w$, where\revisecross{ $\vb{u}=(u,v)$} \revise{$\vb{u}=u\vb{e}_x + v\vb{e}_z$} is the fluid velocity and $\vb{n}$ is a unit vector normal to the rivulet path $\zeta$. The second equation reflects mass conservation.

The LHS of equation~\eqref{eq:ns} represents inertia, where the
numerical prefactor $\beta \simeq 1$ accounts for the velocity's $y$-profile.
Henceforth, following~\cite{daerr2011}, we take $\beta = 1$ to simplify
the equations without losing physical relevance. The terms on the right represent, in this order, gravity, viscous friction internal to the rivulet following from Darcy's law with $\mu = 12\nu/b^2$, streamwise Laplace pressure gradient inside the rivulet due to width variations and forces normal to the rivulet center-line. Note that this streamwise pressure gradient tends to regularize width variations, unlike in the Rayleigh--Plateau instability of cylindrical filaments.

The last term includes three contributions of normal forces. Surface
tension $\gamma$ tends to straighten the rivulet, with $\partial_{xx}\zeta$ being the curvature of the rivulet in the $(x, z)$ plane and \revise{$\Gamma = \pi \gamma/(2\rho)$} \cite{parkHomsy84}. The second term makes for contact line friction, accounting for the high dissipation at the meniscus edges when the rivulet slides transversally on thin films of thickness $h \ll b$ on the plates outside the rivulet: $\mucl \approx \revise{b}\mu\sqrt{b/h}$~\cite[\revisemore{to be published}]{daerr2011}\revise{, with $h = $\SI{3}{\micro\metre} for $Q = $\SI{46 \pm 1}{\cubic\milli\metre\per\second}}.
The last term is the acoustic pressure discussed above.

The base solution is a straight rivulet of constant width $w_0$, located at $z = \zeta_0(t)$ with velocity \revise{$\vb{u}=(u_0\!=\!g/\mu,v_0 = \partial_t\zeta_0(t))$}.\revisemore{ The rivulet rest width $w_0$ in the experimental results presented here was measured to be \SI{0.22\pm 0.04}{\mm} for $Q=\SI{26 \pm 1}{\cubic\milli\metre\per\second}$ and \SI{0.33 \pm 0.04}{\mm} for $Q=\SI{46 \pm 1}{\cubic\milli\metre\per\second}$.}
To understand the mechanism leading to the instability we consider a weak perturbation of this straight rivulet with $u= u_0 + \epsilon u_1(x, t)$, $w = w_0 + \epsilon w_1(x, t)$ and $\zeta = \zeta_0(t) + \epsilon \zeta_1(x, t)$. The transverse speed is $v = (\partial_t + \vb{u}\cdot\nabla)\zeta$. The Navier--Stokes equation~\eqref{eq:ns} projected on $\vb{e}_z$ gives at order~0 
\begin{align}
\label{eq:spaceaverageresponse}
    w_0 (\partial_t + \mu) \partial_t \zeta_0
    = - \mucl \partial_t \zeta_0 + \Pi(Z-\zeta_0) =: w_0 F(t).
\end{align} 
This equation describes the back and forth membrane-like movement of
the rivulet. It is linear in the forcing amplitude, and indeed the
experimental space-averaged rivulet position $\zeta_0$ is always well
fitted by a sine function \revise{of time}. 

At order 1 the same projection yields an equation governing the fluid path $\zeta$,
\begin{align}
    \qty[ w_0 (\partial_t + u_0\partial_x)(\partial_t + u_0\partial_x + \mu) - \Gamma \partial_{xx} + \mucl \partial_t]\zeta_1
    =  - F\, w_1
\label{eq:ns-z-zeta}
\end{align}
which corresponds to equation~(4) from \cite{daerr2011} with an extra
forcing term on the RHS. This term couples the purely \revisecross{temporal global
movement}\revise{time-dependent forcing} $F$ and the width $w$ which is advected at speed $u_0$. It is
destabilizing and causes width modulations in conjunction with
acoustic forcing to amplify path perturbations.

By projecting equation~\eqref{eq:ns} on $\vb{e}_x$, we obtain at first
order in $\epsilon$ the evolution equation for the width $w$,

\begin{align}
    \qty[ (\partial_t + u_0\partial_x)(\partial_t + u_0\partial_x + \mu) + w_0\Gamma \partial_{xxxx}] w_1 &= w_0 F\,\partial_{xx}\zeta_1
\label{eq:ns-x-w}
\end{align}

The right-hand term can be understood as a stretching or compression
of curved rivulet segments by the pressure difference across the
rivulet. This term implies that the growth of width perturbations is
a consequence of path modulations combined with acoustic forcing.

An interesting property of \revise{eqn}.~\eqref{eq:ns-z-zeta}
and~\eqref{eq:ns-x-w} is that the destabilizing RHS does not contain
the quantity differentiated on the LHS. In other words neither the
sinuosity $\zeta_1$ nor the width variations $w_1$ are directly amplified
by the \revisecross{excitation} \revise{forcing}, but rather the acoustic forcing allows mutual
growth by cross-coupling the two modes. While a coupling between
sinuous and width variation modes is not unusual and is found for
example in jets~\cite{MikhaylovWu2020, liebgoldstein89}, usually both
modes are intrinsically unstable, can exist on their own and compete
against each other. Here sinuous and width perturbations are linearly
damped when considered independently, i.e. in the absence of
forcing-induced coupling. They grow only by sustaining one another
through parametric coupling. 

\begin{figure}
    \centering
\includegraphics[width=0.9\maxfigwidth]{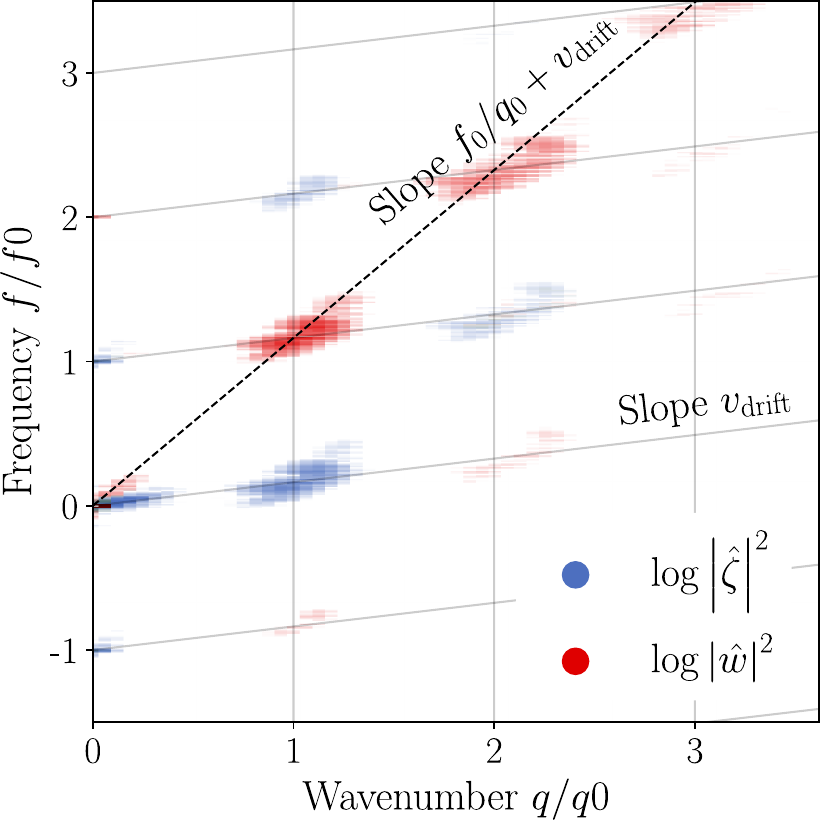}
\caption{Spatio-temporal power spectrum of the position (blue) and
  width (red) of the rivulet for $Q > \Qthres$ ($Q = $\SI{46 \pm 1}{\cubic\milli\metre\per\second}). The frequency scale is set
  by the \revisecross{excitation} \revise{forcing} frequency $f_0 = 50$~Hz, the wavenumber scale by
  the dominant mode $q_0 = 0.193$~mm$^{-1}$. The color intensity
  indicates the strength of the signal on a logarithmic scale, with a
  cut-off to white below -110~dB of the peak signal. An equivalent
  plot for a lower flow rate $Q < \Qthres$ is included in the SI as
  fig.~S1.}
\label{fig:ft}
\end{figure}

We can find experimental confirmation of this mechanism by looking at
the spatio-temporal Fourier transform of the position and width of the
rivulet. Using image analysis tools, we extract from the video the
width and the position of the rivulet as a function of time and the
$x$-coordinate (fig.~\ref{fig:instability} right, fig\revise{s}.~SF2 \revise{and SF3}).
In fig.~\ref{fig:ft} and fig.~S1 we represent the logarithm of
the power spectrum of $\zeta$ and $w$ in the reciprocal space
$(q, f)$. Zones of high intensity are organized in localized spots,
because of the quasi-periodicity of the pattern. For non-zero
wavenumbers, the more intense harmonics are at positive frequencies,
corresponding to positive speed (downward displacement).

More can be said from fig.~\ref{fig:ft} and fig.~S1 on the
spatio-temporal behavior of the rivulet. For instance, note how for
$q=0$, the displacement spectrum $\hat\zeta$ has peaks (blue patches) at
$\pm f_0$ but not at $\pm n f_0$ with $n > 1$: this reflects the fact that
the space-averaged response of the rivulet is linear and the global
movement \revisecross{of the filet} $\zeta_0(t)$ is sinusoidal in time, in
accordance with equation~\eqref{eq:spaceaverageresponse}.

More importantly, the patches of highest power in $\hat w$ (red) lie
on a line whose slope is the speed of the width modulations. The
relevance of the mutual amplification mechanism described above is
supported by the fact that on a given vertical line the high-signal
regions alternate between $\hat w$ (red) and $\hat\zeta$ (blue), the gap
between two consecutive spots being the \revisecross{excitation} \revise{forcing} frequency $f_0$.
This shows that both modes are coupled by a function oscillating at
$f_0$, and that no mode is unstable on its own (or we would see all
its time harmonics). Finally the localization of the spectrum at
discrete wavenumbers is also explained by our model, positive feedback
and mutual amplification of longitudinal and transverse waves requiring
the resonance condition $u_0\,q = \vdrift\,q + n\,f_0$ to be met.


\paragraph{Conclusion and perspectives} Under the effect of a
spatially uniform forcing, a homogeneous membrane or string is
expected to respond by homogeneous transverse translation. Remarkably,
the liquid rivulet studied in this letter develops a pattern with a
well defined wavelength combining transverse deformations of the flow
path and longitudinal modulations of the local cross-section.

It is far from obvious that the added degree of freedom with respect
to a string, namely the possibility to redistribute mass along the
rivulet, should lead to an instability. Indeed both modes growing
simultaneously, sinuosity and width variations, are linearly damped,
and they amplify one another only when coupled by the forcing. This
contrasts with vibrated soap films and strings loaded with beads,
where mass redistribution merely causes the broadening of
resonances\cite{boudaoud_bead, boudaoud99}.



Interestingly, although the acoustic forcing is additive, the effective forcing felt by the sinuosity and width variations is multiplicative. Formally this parametric cross-coupling is reminiscent of the way standing waves in the annular Faraday instability can be seen as arising through the coupling of counter-propagating waves~\cite{douady1989}. \revise{The resonance condition to the amplification indicates a possible connection with the energy transfer due to resonant three-wave interaction in the case of stratified or homogeneous free-surface flows over a non-flat bottom \cite{Benilov1987, Szoeke1983}.}

The selection of the pattern drift speed is an open problem. We attributed modifications of the width advection speed at high frequencies to asymmetric deformations of the rivulet cross-section. The study of these deformations, which can lead to rivulet breaking, could open the perspective of investigating the problem of the behavior of an air-fluid interface in a Hele-Shaw cell in the oscillating regime where inertial effects can be as important as capillary ones\cite{aussillous2000}. The forced rivulet also allows the coupling and simultaneous study of both the retreating and advancing menisci. 

As an experimental investigation of a previously unreported instability, validated by theoretical modeling that identifies the mechanism as a new type of parametric coupling, this Letter opens exciting perspectives for new research and applications. The generation of wavelengths that are orders of magnitude smaller than the acoustic wavelength of the forcing could be exploited for controlled liquid fragmentation, mixing, and micro-manufacturing. We also expect fundamental research on dynamical wetting to take advantage of the broad frequency response, that can for instance be used to probe timescales relevant in surfactant diffusion/adsorption/desorption dynamics.

\paragraph{Acknowledgements} We thank Michael Berhanu and Chi-Tuong Pham for insightful discussions. The research received special funds from laboratory MSC UMR 7057. For the purpose of Open Access, a CC-BY public copyright licence has been applied by the authors to the present document and will be applied to all subsequent versions up to the Author Accepted Manuscript arising from this submission.


\providecommand{\noopsort}[1]{}\providecommand{\singleletter}[1]{#1}%

\end{document}